\documentclass[11pt]{article}
\input{style.sty}

\title{Lorentzian Path Integrals and Jackiw-Teitelboim wormholes with imaginary scalars}
\author[a]{Jesse Held,} \emailAdd{jheld@ucsb.edu}
\author[b]{Molly Kaplan,} \emailAdd{molly.kaplan@inria.fr}
\author[a]{Donald Marolf,} \emailAdd{marolf@ucsb.edu}
\author[c]{Zhencheng Wang} \emailAdd{zcwang1@illinois.edu}

\affiliation[a]{Department of Physics, University of California, Santa Barbara, CA 93106, USA}
\affiliation[b]{Laboratoire de Physique de l’École Normale Supérieure, Mines Paris, Inria, CNRS, ENS-PSL, Centre Automatique et Systèmes (CAS),
Sorbonne Université, PSL Research University, Paris, France}
\affiliation[c]{Department of Physics, University of Illinois Urbana-Champaign, Urbana, IL 61801, USA}

\abstract{The Lorentzian path integral was recently used to argue that standard Euclidean axion wormholes do not dominate computations of connected AdS/CFT partition functions.  We now apply similar methods to study the seemingly-analogous Jackiw-Teitelboim wormholes constructed by Garcia-Garcia and Godet  using Jackiw-Teitelboim gravity with an imaginary-valued minimally-coupled massless scalar field.  However, this time we find that these wormholes {\it do} dominate our path integral for the relevant connected partition function.  This supports the suggestion by Garcia-Garcia and Godet that contributions from such wormholes parallel the physics of the Sachdev-Ye-Kitaev model at complex couplings.  The result also illustrates the sensitivity of wormhole contributions to details of the relevant physics.}

\begin{document}
\maketitle
\section{Introduction}

It is well-known  that contributions to gravitational path integrals from spacetime wormholes can lead to S-matrices, boundary correlators, and boundary partition functions that fail to factorize over disconnected boundaries \cite{Lavrelashvili:1987jg,Hawking:1987mz,Hawking:1988ae,Coleman:1988cy,Giddings:Original_Axion_Wormhole,Giddings:1988cx}.  A particularly-interesting such wormhole was constructed in \cite{Garcia-Garcia:2020ttf} (see also \cite{Garcia-Garcia:2022zmo}) by coupling Jackiw-Teitelboim (JT) gravity to an imaginary-valued massless-scalar.  Contributions from this wormhole were argued in those works to describe the same physics one finds in the Sachdev-Ye-Kitaev (SYK) model with complex couplings; see also \cite{Maldacena:2018lmt}.

The Euclidean wormhole of \cite{Garcia-Garcia:2020ttf} is qualitatively similar to AdS-axion wormholes that one finds in higher dimensions (see e.g. \cite{Arkani-Hamed:2007cpn}, building on the asymptotically flat solutions of \cite{Giddings:Original_Axion_Wormhole}).  Both involve a massless scalar (or pseudo-scalar) that, from the perspective of standard Lorentzian physics, takes imaginary values and so has a `wrong-sign' kinetic term -- though in the axion case this is often said to be a consequence of Wick-rotating a pseudoscalar to Euclidean signature; see further related comments in \cite{AxionWormholes}.  In both cases, it is the associated wrong-sign kinetic energy that supports the wormhole.

However, it was recently argued \cite{AxionWormholes} that AdS-axion wormholes provide at most a subdominant contribution to the computation of connected boundary partition functions of the sort considered in \cite{Garcia-Garcia:2020ttf,Garcia-Garcia:2022zmo}.  Here we use the term `at most' to indicate that the saddle in fact lies on a Stokes' line across which the appropriate ascent contour changes from having intersection number zero with the contour of integration (signifying that the saddle does not meaningfully contribute to the integral) to having intersection number one -- though even on the latter side the saddle remains subdominant.

The argument in \cite{AxionWormholes} was based on resolving the conformal-factor problem of Euclidean gravity \cite{Gibbons:1978ac} by instead defining the path integral using a real Lorentzian contour.  This basic idea has been advocated by many authors \cite{Hartle:2020glw,Schleich:1987fm,Mazur:1989by,Giddings:1989ny,Giddings:1990yj,Marolf:1996gb,Dasgupta:2001ue,Ambjorn:2002gr,Feldbrugge:2017kzv,Feldbrugge:2017fcc,Feldbrugge:2017mbc,Brown:2017wpl,Marolf:2020rpm,Colin-Ellerin:2020mva,Colin-Ellerin:2021jev, Marolf:Thermodynamics}, though the details of the approach used in \cite{AxionWormholes} were formulated more recently in \cite{Marolf:2020rpm,Colin-Ellerin:2020mva,Colin-Ellerin:2021jev,Marolf:Thermodynamics}.

Ref. \cite{AxionWormholes} then evaluated the resulting Lorentzian path integrals by noting that, if one holds fixed the cross-sectional area of the wormhole at the locus where the axion vanishes, one can then find real Lorentz-signature wormhole saddles for the corresponding constrained variational problem\footnote{In contrast, without this constraint real Lorentz-siganture saddles are forbidden by the topological censorship theorems \cite{Friedman:1993ty,Galloway:1999bp}.}. After arguing that these constrained wormholes could be used to evaluate this constrained path integral in the semiclassical approximation, the remaining integrals over the above area and a time-period parameter $T$ (and third parameter associated with a Lorentzian analogue of a conical defect) were then studied in detail to argue that the axion wormhole was at best marginally-relevant (and subdominant) at the associated boundary-values of the axion.  

The apparent similarity between the imaginary scalar JT wormhole and higher-dimensional axion wormholes thus leads to a tension between the proposal of \cite{Marolf:2020rpm,Colin-Ellerin:2020mva,Colin-Ellerin:2021jev,Marolf:Thermodynamics} and the match described in \cite{Garcia-Garcia:2020ttf,Garcia-Garcia:2022zmo} between SYK physics and that of the JT wormhole with imaginary scalar.  
We investigate this tension here by directly applying the methods of \cite{Marolf:2020rpm,Colin-Ellerin:2020mva,Colin-Ellerin:2021jev,Marolf:Thermodynamics} to  gravity with a minimally-coupled massless scalar.

In particular, 
as in \cite{AxionWormholes}, we impose Dirichlet boundary conditions on the scalar, fixing the values at the left and right boundaries to some $\phi_L,\phi_R$.  
In parallel with \cite{AxionWormholes}, we first formulate the problem for real $\phi_L,\phi_R$, and then analytically continue the formalism to imaginary $\phi_L,\phi_R$.  We also follow \cite{AxionWormholes} by first fixing a cross-sectional area $A_0$ and a time-period parameter $T$, evaluating the associated fixed-area path integral using the saddle-point approximation with a fixed-area saddle that lies on the original (real and Lorentz-signature) contour integration, and then studying the remaining $A_0,T$ integrals in detail.  

However, the results for the JT-scalar system turn out to be very different from those found in \cite{AxionWormholes}.  While the wormhole saddle still does not contribute for real $\phi_L,\phi_R$, the Stokes' line is crossed much earlier and, by the time we analytically continue to imaginary $\phi_L,\phi_R$ the Euclidean wormhole does in fact dominate the computation.  This appears to fully resolve the above-noted tension.

We begin in section \ref{sec:PILorentz} below by very briefly reviewing the way in which \cite{Marolf:Thermodynamics} formulates thermal partition functions in terms of Lorentzian path integrals.  We then set up the desired JT-gravity computation by reviewing the construction from \cite{Garcia-Garcia:2020ttf} of Euclidean JT wormholes in section \ref{sec:review}.   Lorentzian wormhole constrained-saddles case are then constructed and used to analyze the desired path integral in section \ref{sec:LorentzJT} for general complex boundary-values of the scalar field.  We close with a brief summary and discussion of open issues in section \ref{sec:disc}.

\section{Partition functions from Lorentzian Path Integrals}
\label{sec:PILorentz}

This section provides an extremely brief review of the procedure from \cite{Marolf:Thermodynamics} for formulating thermal partition functions in terms of Lorentzian path integrals.  Interested readers are encouraged to consult the longer review in \cite{AxionWormholes}, or to consult the original work.  However, 
many of the details discussed in those references turn out to be irrelevant to the JT wormholes we analyze in this work.  This simplification stems from the fact that JT gravity is two-dimensional, so that any closed boundary is a union of circles.   Our JT wormholes will thus have topology $S^1 \times {\mathbb R}$.  In contrast to the cases of most interest in \cite{Marolf:Thermodynamics} and \cite{AxionWormholes}, smooth Lorentz-signature metrics on $S^1 \times {\mathbb R}$ are perfectly compatible with taking the boundary circles to be timelike.  As a result, we will have no reason to introduce the singularities in the Lorentzian structure that were needed in the cases of most interest in 
\cite{Marolf:Thermodynamics,AxionWormholes} and which required more detailed treatment.

Before addressing the gravitational case, we begin by considering the partition function $Z(\beta) = \Tr  e^{-\beta H}$ in a non-gravitating quantum system.  Here $H$ is a Hamiltonian that we assume to be bounded below by some  energy $E_0$. So long as $E_0$ is slightly below the actual ground state energy, we may freely and unambiguously replace  $e^{-\beta H}$ with the product $e^{-\beta H}\theta(H-E_0)$.  Here, as usual, $\theta$ denotes the Heaviside step function that satisfies $\theta(x) =1$ for $x< 0$ and $\theta(x) = 0$ for $x < 0$.  

Noting that the function $e^{-\beta E}\theta(E-E_0)$ is square-integrable, we may use its appropriately-normalized Fourier transform $f_\beta(T)$ to write the above operator as 
\begin{equation}
\label{eq:expFT}
e^{-\beta H}  =\theta(H-E_0)e^{-\beta H} = \int_{-\infty}^{\infty} dT \, f_\beta(T) e^{-iHT}.
\end{equation}
With the above conventions, the relevant $f_\beta(T)$ is in fact just
        \begin{equation}
            f_\beta(T)= \frac{e^{iTE_0-\beta E_0}}{2\pi i(T+i\beta)},
        \end{equation}
which features a pole at $T=-i\beta$.  Indeed, since $H \ge E_0$, the factor of $iTE_0$ in $f_\beta$ allows one to close the contour of integration for \eqref{eq:expFT} in the 
lower-half of the complex $T$-plane.  It is straightforward to obtain the left-hand side of \eqref{eq:expFT}  by using the residue at $T=-i\beta$ to  evaluate the right-hand-side.  
Taking the trace of \eqref{eq:expFT} then yields
\begin{equation}
\label{eq:ZFT}
Z(\beta) = \Tr \int_{-\infty}^{\infty} dT \, f_\beta(T) e^{-iHT}.
\end{equation}

All that remains is to understand is the gravitational analogue of \eqref{eq:ZFT}.  The proposal of \cite{Marolf:Thermodynamics} is motivated by supposing that one can interchange the trace and the $T$-integral in \eqref{eq:ZFT} to write $Z(\beta)$ in terms of $\Tr e^{-iHT}$.  This latter object would then naturally be given by an integral over real Lorentz-signature bulk spacetimes for which time is periodic (with period $T$) on any boundaries. The proposal of \cite{Marolf:Thermodynamics} is thus to write
\begin{equation}
\label{eq:Zgrav}
Z(\beta) = \int {\cal D}g {\cal D}\phi \  f_\beta(T) \ e^{iS_{bulk}}.
\end{equation}
Here $\phi$ denotes an arbitrary collection of matter fields and \eqref{eq:Zgrav} integrates  over Lorentz-signature spacetime metrics $g$ with periodic boundary time, and where $T$ is now just the name for this period.  In particular, in the asymptotically AdS case of interest here, we assume that we are given a one-parameter family of static time-periodic boundary metrics $B_T$ where $T\in {\mathbb R}$ denotes the period of the relevant notion of boundary time and where, as we will see, $T\rightarrow -T$ corresponds to changing the sign of the real part of the Lorentzian action.  Our notation then includes this T as one of the degrees of freedom of the bulk metric $g$.  As noted above, in higher dimensions the periodicity of our boundary time would require singularities in the bulk causal structre that would require further prescriptions.  However, this will not be required in the context studied in this work.

Despite the above motivations, in analogy with the lack of absolute convergence one finds if one tries to perform the trace in \eqref{eq:expFT} before integrating over $T$, one suspects it to be useful in \eqref{eq:Zgrav} to save the integrations of at least some class of metric degrees of freedom until after the $T$-integral has been performed.  Let us denote the relevant class of metric variables by the collective label $\nu.$  We will thus in fact write
\begin{equation}
\label{eq:Zgrav2}
Z(\beta) = \int d\nu \int_{-\infty}^\infty dT   \  f_\beta(T)\  Z(B_T;\nu) ,
\end{equation}
with the $T$-integral being performed before the $\nu$-integral as described above.  Here the constrained partition function $Z(B_T;\nu)$ is given by the constrained path integral 
\begin{equation}
\label{eq:ZgravTnu}
Z(B_T;\nu) = \int_{\partial M = B_T; \nu} {\cal D}g {\cal D}\phi  \ e^{iS_{bulk}},
\end{equation}
where the notation indicates that we now sum only over bulk spacetimes with the given values of the collective parameter $\nu$.  We will find the objects $Z( B_T;\nu)$ to be be well-defined at least at leading semiclassical order. 

Finally, in using this formalism to study spacetime wormholes, we will be interested in situations where each $B_T$ has two connected components, $B_T = B_T^L \sqcup B_T^R$, corresponding to the left and right boundaries of the wormhole.  In this context we will use the notation
\begin{eqnarray}
\label{eq:ZgravNot}
Z_{L,R}(\beta) = \int d\nu \int_{-\infty}^\infty dT   \  f_\beta(T)\  Z(B_T^{L,R};\nu), \ &\ &\ 
Z_{L \sqcup R}(\beta) = \int d\nu \int_{-\infty}^\infty dT   \  f_\beta(T)\  Z(B_T^L\sqcup B_T^R;\nu), \\
&{\rm and }&\text{ } Z^C_{L \sqcup R}(\beta) := Z_{L \sqcup R}(\beta) - Z_{L}(\beta)Z_{R}(\beta).
\end{eqnarray}
We will refer to $Z^C_{L \sqcup R}(\beta)$ as a connected partition function.

\section{Review: Euclidean wormholes in JT gravity}
\label{sec:review}

An interesting class of wormhole solutions in Euclidean Jackiw-Teitelboim (JT) gravity coupled to a massless scalar field were discussed in \cite{Garcia-Garcia:2020ttf}.  Such Euclidean wormholes connect two asymptotically AdS boundaries, which are each topologically just a circle.  The authors of \cite{Garcia-Garcia:2020ttf} argue that such wormholes lead to important effects, since the contribution of such wormholes can be qualitatively matched to a corresponding calculation in complex SYK model. 

We now review the wormhole solutions of \cite{Garcia-Garcia:2020ttf}. The Euclidean action for JT gravity with a (minimally coupled) massless scalar field $\chi$ is
\begin{equation}
    I=I_{\mathrm{JT}} + I_{\mathrm{matter}},
\end{equation}
where
\begin{equation}
    I_{\mathrm{JT}}=-\frac{S_0}{2 \pi}\left[\frac{1}{2} \int_{\cal M} d^2 x \sqrt{g} R+\int_{\partial \cal M} d \tau \sqrt{h} K\right]-\frac{1}{2} \int_{\cal M} d^2 x \sqrt{g} \Phi(R+2)-\int_{\partial \cal M} d \tau \sqrt{h} \Phi(K-1),
\end{equation}
\begin{equation}
{\rm and} \ \     I_{\text {matter }}=\frac{1}{2} \int_{\cal M} d^2 x \sqrt{g}(\partial_\mu \chi \partial^\mu\chi).
\end{equation}
Varying the action yields the equations of motion
\begin{equation}
   \square \chi = 0,\quad R=-2,\quad \nabla_\mu \nabla_\nu \Phi-g_{\mu \nu} \square \Phi+g_{\mu \nu} \Phi+T_{\mu \nu}^\chi=0 
\end{equation}
where the stress tensor of the scalar field is\footnote{In \cite{Garcia-Garcia:2020ttf}, the authors studied the expectation value $\la T_{\mu \nu}\ra$ of the scalar field quantum stress tensor, finding a quantum correction term that plays an important role at low temperatures. However, a purely classical treatment will suffice for our purposes below.}
\begin{equation}
    T_{\mu\nu}^\chi =  \partial_\mu\chi \partial_\nu\chi - \frac{1}{2}g_{\mu \nu} \partial_\rho \chi \partial^\rho \chi.   
\end{equation}

As in \cite{Garcia-Garcia:2020ttf}, we will focus on wormhole solutions with a $U(1)$ rotational symmetry. We find it convenient for later purposes to use the following metric ansatz (which differs by a change of coordinates from the one used in \cite{Garcia-Garcia:2020ttf}).  In particular, we take
\begin{equation}
    ds^2=dr^2+f(r)d\tau^2,\quad \tau\sim \tau+\beta
\end{equation}
where $r=\pm \infty$ correspond to two asymptotic boundaries.

With this metric ansatz, the above equations of motion become
\begin{equation}
\label{eq:scalarEOM}
    \chi''(r)+\frac{f'(r)}{2f(r)}\chi'(r)=0,
\end{equation}
\begin{equation}
\label{eq:R=-2}
    R= \frac{f'(r)^2-2f(r)f''(r)}{2f(r)^2}=-2,
\end{equation}
\begin{equation}
\label{eq:PhiEoM}
    \Phi''(r) -\Phi(r)+\frac{1}{2}\chi'(r)^2=0,\quad 
    \Phi(r)-\frac{f'(r)\Phi'(r)}{2f(r)}+\frac{1}{2}\chi'(r)^2=0.
\end{equation}

In \cite{Garcia-Garcia:2020ttf}, the authors found that wormhole solutions exist when the scalar field takes purely imaginary values. In particular, there is a one-parameter family of $\mathbb{Z}_2$ symmetric wormhole solutions labeled by $k_E\in {\mathbb R}$ and given by\footnote{Ref. \cite{Garcia-Garcia:2020ttf} also presents smooth wormhole solutions where both the wormhole and the boundary conditions break this $\mathbb{Z}_2$-symmetry.  However, for simplicity we study only the $\mathbb{Z}_2$-symmetric ones here.}
\begin{equation}
\label{eq:JTEuclideanWormhole}
\begin{aligned}
    \chi(r)&=\frac{4ik_E}{\pi} \tan^{-1}\left(\tanh \frac{r}{2}\right)\\
    f(r) &= \frac{k^4_E}{\pi^2 \phi_b^2}\cosh^2 r\\
    \Phi(r) &= \frac{2k^2_E}{\pi^2} \left(1+2\tan^{-1} (\tanh \frac{r}{2}) \sinh r \right).
\end{aligned}
\end{equation}
This solution satisfies the boundary conditions
\begin{equation}
    \chi(\pm \infty) = \pm i k_E,\quad \Phi|_{\mathrm{bdy}}=\frac{\phi_b}{\epsilon} +O(\epsilon^0), \quad  {\rm and} \quad \int \sqrt{h}|_{\mathrm{bdy}}=\frac{\beta}{\epsilon} +O(\epsilon^0),
\end{equation}
where the notation $|_{\mathrm{bdy}}$ means that the relevant field is evaluated on cutoff surfaces $r=\pm \ln \frac{2\pi\phi_b}{k^2_E\epsilon}$ in the limit $\epsilon \rightarrow 0$.

It is not hard to compute the Euclidean action of this solution. Since $R=-2$ everywhere, we only need to evaluate the boundary term and the action of the scalar field. The on-shell action is thus
\begin{equation}
    I=-\int_{\partial \cal M} d \tau \sqrt{h} \Phi(K-1)+\frac{1}{2} \int_{\cal M} d^2 x \sqrt{g}(\partial_\mu \chi \partial^\mu\chi) = - \frac{k^4_E\beta}{\pi^2 \phi_b}.
\end{equation}

\section{Lorentzian gravitational path integral for wormholes in JT gravity}
\label{sec:LorentzJT}

We wish to determine whether the saddle reviewed above contributes to the gravitational path integral for the partition function $Z_{L\sqcup R}(\beta)$ defined in \eqref{eq:ZgravNot}.   Since we work in 1+1 dimensions, we take both $B_T^L$ and $B_T^R$ to be timelike circles with circumference of proper time $|T|$.  As in \cite{AxionWormholes}, we will choose the constraints $\nu$ so that they define a constrained variational problem that has Lorentz-signature wormhole solutions. This is done in section \ref{sec:WCW}. We will then study the integral over the relevant constrained wormholes in detail in section \ref{sec:int}.

\subsection{Wormholes and constrained wormholes}
\label{sec:WCW}

In Lorentzian signature, the action for JT gravity minimally coupled to scalar field is given by
\begin{equation}
    S=S_{\mathrm{JT}} + S_{\text {matter }}
\end{equation}
where
\begin{equation}
    S_{\mathrm{JT}}=\frac{S_0}{2 \pi}\left[\frac{1}{2} \int_{\cal M} d^2 x \sqrt{-g} R+\int_{\cal \partial M} d \tau \sqrt{-h} K\right]+\frac{1}{2} \int_{\cal M} d^2 x \sqrt{-g} \Phi(R+2)+\int_{\cal \partial M} d \tau \sqrt{-h} \Phi(K-1),
\end{equation}
\begin{equation}
    S_{\text {matter }}=-\frac{1}{2} \int_{\cal M} d^2 x \sqrt{-g}(\partial_\mu \chi \partial^\mu\chi).
\end{equation}
To find wormhole solutions, we will use the metric ansatz 
\begin{equation}
\label{eq:LorentzianMetric}
    ds^2=dr^2-f(r)dt^2
\end{equation}
and impose boundary conditions analogous to those in section \ref{sec:review}:
\begin{equation}
\label{eq:JTBC}
\chi(\pm\infty)=\pm k, \quad
\phi|_{\mathrm{bdy}}=\frac{\phi_b}{\epsilon} ,\quad \int_{\mathrm{bdy}}\sqrt{-h}=\frac{T}{\epsilon} \quad \text{at}\quad r=\log \frac{2\pi \phi_b}{k^2 \epsilon}
\end{equation}
with $\phi_b>0$ and $k\in\mathbb{R}$.  We also take $t$ to have period $|T|$. 

Let us first discuss possible wormhole solutions for JT with minimally coupled scalar in Lorentzian signature. When the scalar field is real, we can construct formal solutions by analytically continuing the Euclidean wormhole \eqref{eq:JTEuclideanWormhole} according to  $\tau \to i t$ and $k_E\to -ik$. However, this makes the dilaton negative at the asymptotic boundaries, forcing $\phi_b<0$ and violating the boundary conditions \eqref{eq:JTBC}. Indeed, by the JT analogue of topological censorship, there can be no real wormhole solutions in Lorentz signature.

However, one would expect that we {\it can} construct constrained wormhole saddles with appropriate constraints.  Natural analogues of the constraint imposed in \cite{AxionWormholes} would be to fix either the metric to have $f^{1/2}(r_0)=A_0/T$ at the value of $r$ where $\chi=0$, or to fix the dilaton to $\Phi=\Phi_0$ at that locus. 
The first parameter $A_0$ is named in reference to the fact that, in higher dimensions, $A: = f^{1/2}T$ would correspond to an appropriate (signed\footnote{Specifically, with a sign chosen to match the sign on $T$.}) notion of cross-sectional area for the wormhole at each value of $r$.  Since our constrained wormholes will again have $U(1)$ symmetry, for convenience we refer to this locus as the surface $r=r_0$.

We note, however, that  equation of motion $R=-2$ is already compatible with the existence of real Lorentz-signature wormholes as can be seen by identifying global AdS$_{1+1}$ under a global-time translation.  Indeed, we recall that it is the dilaton equation of motion that forbids such Lorentz-signature wormholes\footnote{This can be seen from the fact that the dilaton along a null line is the analogue of horizon area in higher dimensions, and that its equation of motion thus provides a JT analogue of the Raychaudhuri equation.}.  Allowing real Lorentz-signature wormholes thus requires the addition of new sources to the dilaton equation of motion.  Furthermore,  since (the derivative of) the dilaton is canonically conjugate to the metric, such sources are added by imposing constraints on the metric rather than fixing $\Phi$ at $r_0$.  For this reason, we restrict attention below to imposing only the constraint $f^{1/2}(r_0)=A_0/T$.

The associated constrained variational principle can now be defined by adding appropriate Lagrange multipliers to the action.  In particular, we take
\begin{equation}
\label{eq:SwLMs}
    S_{\mathrm{JT}}+S_{\mathrm{matter}} 
    + \lambda_A \left[\left( \int d^2x\, \sqrt{g}r^\mu\nabla_\mu \sgn(\chi)\right) -A_0\right],
\end{equation}
 where $r^\mu$ is the unit normal to the $\chi=0$ surface directed toward the right boundary and where we have taken care to write our area-constraint in a form that is manifestly invariant under diffeomorphisms and which does not yet assume the desired $U(1)$ symmetry.
This action results in the following constrained equations of motion:
\begin{align}
    R+2
    =0\\
    \quad \nabla_\mu \nabla_\nu \Phi-g_{\mu \nu} \square \Phi+g_{\mu \nu} \Phi+T_{\mu \nu}^\chi
    +\lambda_A \delta(r-r_0) =0\\
    A|_{r=r_0}=A_0. 
\end{align}

We will now seek solutions to the above equations consistent with the boundary conditions \eqref{eq:JTBC}. For simplicity, we will restrict attention to constrained wormholes with $\mathbb Z_2$ symmetry across $r=r_0$.  It is therefore sufficient to only study the `half-wormholes' with $r\in [r_0, \infty]$ and to then paste together two copies of the half-wormhole to construct our $\mathbb Z_2$-symmetric wormhole. 

Let us first find the most general $U(1)$-symmetric half-wormhole solution consistent with our metric ansatz \eqref{eq:LorentzianMetric}. Due to the $U(1)$ symmetry, the equations of motion take the same form as in Euclidean signature; i.e., they are again given by \eqref{eq:scalarEOM}, \eqref{eq:R=-2} and \eqref{eq:PhiEoM}. Requiring $R=-2$ \eqref{eq:R=-2} forces $f(r)$ in the metric to take the form 
\begin{equation}
\label{eq:generalf}  f(r)=c_1^2(e^r+c_2^2e^{-r})^2,
\end{equation}
where $c_1$ and $c_2$ are undetermined constants. From \eqref{eq:scalarEOM}, we find $\chi(r)=\int \frac{1}{\sqrt{f(r)}}dr$, and then by using \eqref{eq:generalf} and enforcing the scalar boundary conditions $\chi(r_0)=0$ and $\chi(+\infty)=k$ we find
\begin{equation}
    \chi(r)=\frac{2k(R(r)-R_0)}{\pi-2R_0},
\end{equation}
where
\begin{equation}
R(r)\equiv \tan^{-1}(e^{r}/c_2),\quad R_0\equiv R(r_0).
\end{equation}
Finally, solving the dilaton equations of motion \eqref{eq:PhiEoM} using the above $f(r)$ and $\chi(r)$ gives
\begin{equation}
    \Phi(r)=-\frac{k^2 c_2+ k^2 (e^r -c_2^2 e^{-r})R(r)}{2c_2 (\pi-2R(r_0))^2} - c_3(e^r - c_2^2 e^{-r}).
\end{equation}
where $c_3$ is an additional undetermined constant.

Imposing the boundary conditions \eqref{eq:JTBC} fixed two of these constants, requiring 
\begin{equation}
    c_1= \frac{k^2}{2\pi},\quad {\rm and} \quad c_3 = -\frac{k^2}{4\pi}\left( \frac{\pi^2}{c_2(\pi-2R_0)^2} +2\right).
\end{equation}
We may also use the constraint $A|_{r=r_0}=A_0$ to fix $c_2$ to be 
\begin{equation}
    c_2 = -\frac{2\pi A_0 e^{r_0}\phi_b}{k^2 T} - e^{2r_0}.
\end{equation}
Finally, we note that the equation $R=-2$ at $r=r_0$  requires the metric to be smooth there.  The $\mathbb Z_2$-symmetry thus imposes  $f'(r_0)=0$.  This fixes the last undetermined constant $r_0$ to satisfy
\begin{equation}
    e^{r_0}=\frac{A_0 \pi \phi_b}{k^2 T}.
\end{equation}

For future reference, we note that the smooth Euclidean wormhole solution has negative $f$ with
\begin{equation}
    c_1= i\frac{k^2\beta}{2\pi\phi_b},\quad c_2=1,\quad c_3=\frac{k^2}{2\pi},\quad r_0=0
\end{equation}
and, as a result, $A_0=i\frac{k^2\beta}{\pi \phi_b}$.

\subsection{Action and Partition Functions for real boundary conditions}
\label{sec:int}

We can now compute the gravitational action for our constrained wormhole and use it to evaluate the connected partition function as described in section \ref{sec:PILorentz}.  We will begin by discussing the case of real $k=\chi(+\infty)$ and then later analytically continue our analysis to imaginary $k$.  

Since  the dilaton is continuous at $r=r_0$, $R=-2$ everywhere,  and the Lagrange multiplier terms in \eqref{eq:SwLMs} vanish when the constraints are satisfied, non-vanishing contributions to the action arise only from the boundary terms at the asymptotic boundaries,
\begin{equation}
    S_{JT}=2\times \left(  \int_{r=\infty} d\tau\,\sqrt{-h} \Phi (K-1) \right)=-\frac{A_0^2}{T}
\end{equation}
and the matter action
\begin{equation}
    S_{matter}=-2\times \left(\frac{1}{2} \int d\tau \int_{r_0}^{+\infty} dr\, \sqrt{-g}g^{rr}(\partial_r \chi)^2\right)=-\frac{2k^2A_0}{\pi},
\end{equation}
where the factors of 2 come from the two halves of the wormhole. The total action for the constrained wormhole is then
\begin{equation}
    S_{CWH}= S_{JT}+S_{matter}=- \frac{ \phi_b}{T}A_0^2-\frac{2k^2}{\pi}A_0.
\end{equation}

Using these wormholes to evaluate $Z^C(B_T^L\sqcup B_T^R;\nu): = Z(B_T^L\sqcup B_T^R;\nu)-Z^C(B_T^L;\nu)Z^C(B_T^R;\nu)$ at leading order in the semiclassical approximation\footnote{\label{foot:definingC} Since real Lorentz-signature constrained wormhole saddles at fixed positive real $A_0,T$ lie on the original contour of integration, and since the integrand is a pure phase when evaluated on that contour, such constrained saddles necessarily contribute to our constrained path integral.  See e.g. section 3 of \cite{Marolf:Thermodynamics} or appendix A of \cite{AxionWormholes}.}, we can use \eqref{eq:ZgravNot} to write the desired $Z^c_{L\sqcup R}(\beta)$ as an integral over two variables:
\begin{equation}
    Z^C_{L \sqcup R}(\beta) \approx \int_{\frac{A_0}{T} \ge 0} dT dA_0 \,f_\beta(T) e^{iS_{CWH}(A_0,T)} = \int_{\tilde A_0 \ge 0} dT d\tilde A_0 f_\beta(T) \exp\left( -i \phi_b  \tilde A_0^2 T -\frac{2ik^2}{\pi} \tilde A_0 T \right),
\end{equation}
 where we have defined $\tilde A_0\equiv A_0/T$.  Note that $\tilde A_0$ is always positive since we defined the sign of $A_0$ to agree with the sign of $T$. The integration contour on the right-hand side thus has $\tilde A_0 \in [0,+\infty)$. Following \cite{Marolf:Thermodynamics,AxionWormholes}, we first perform the $T$-integral by 
noting that $(\phi_b \tilde A_0^2+2k^2 \tilde A_0/\pi)$ is positive (since both terms are positive separately) so that we may close  the $T$-contour in the lower half-plane.  We then express the integral in terms of the residue of the pole at $T=-i\beta$. Doing so yields
\begin{equation}
\label{eq:LZ}
    Z^C_{L \sqcup R}(\beta) = \int_{\tilde A_0 \in {\mathbb R}^+} d\tilde A_0 \, \exp \left( - \phi_b \tilde A_0^2\beta  - \frac{2k^2}{\pi} \tilde A_0\beta \right).
\end{equation}
The integrand is Gaussian in $\tilde A_0$, though the saddle is located at $\tilde A_0=-\frac{k^2}{\pi \phi_b}$.  Since this value is negative, it does {\it not} lie on the integration contour.  The fact that the integrand vanishes as $\tilde A_0 \to +\infty$ thus implies that the integral is dominated by the endpoint contribution at $\tilde A_0=0$.  Using \eqref{eq:LZ} thus yields  $Z^C_{L \sqcup R}(\beta) = O(1)$ at leading order in the semiclassical approximation $\beta \phi_b \gg 1$.

\subsection{Analytic continuation to general $k\in {\mathbb C}$}
\label{sec:ack}

We now study the analytic continuation of the partition function to more general $k\in {\mathbb C}$.  We emphasize that, since the analytic continuation of small quantities can be large, this cannot be done by studying our final result that, for real $k$, we have $Z^C_{L \sqcup R}(\beta) = O(1)$.  Instead, we must reconsider each stage in the calculation above.

Now, the constrained wormholes above certainly exist for general complex values of $k$.  It is thus reasonable to suppose that they continue to control the computation of $Z^C(B_T^L\sqcup B_T^R;\nu)$ for general complex $k$.  Making that assumption then leads us to study the analytic continuation of \eqref{eq:LZ}.  This is straightforward due to the fact that the coefficient of $\tilde A_0^2$ is negative and completely independent of $k$.  Thus the integral over $\tilde A_0 \in {\mathbb R}^+$ converges for all complex $k$, and the usual argument shows that the result satisfies the Cauchy-Riemann equations.  It thus gives the desired analytic function of $k$.

\begin{figure}
\centering
\begin{minipage}{0.375\textwidth}
        \includegraphics[width=0.8\linewidth]{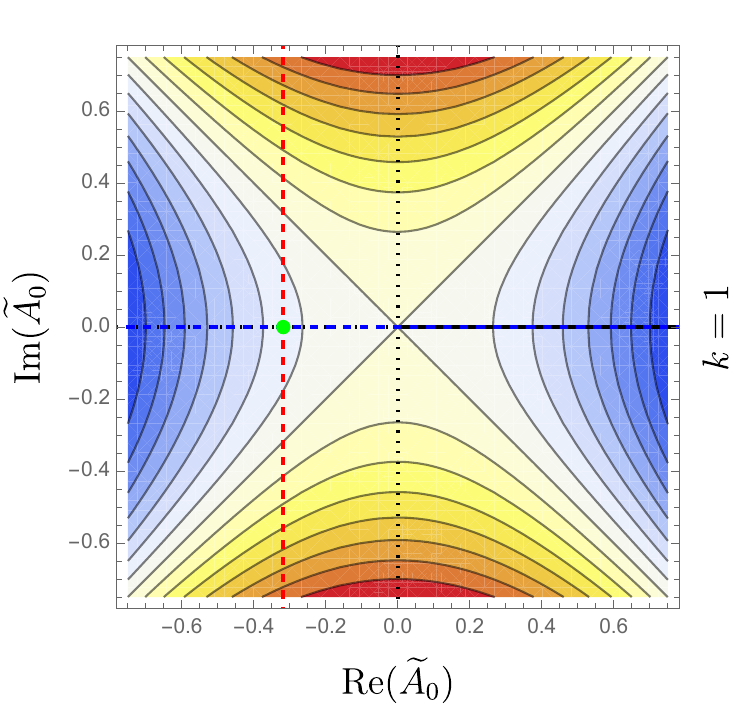}
\end{minipage}
\begin{minipage}{0.375\textwidth}
        \includegraphics[width=0.8\linewidth]{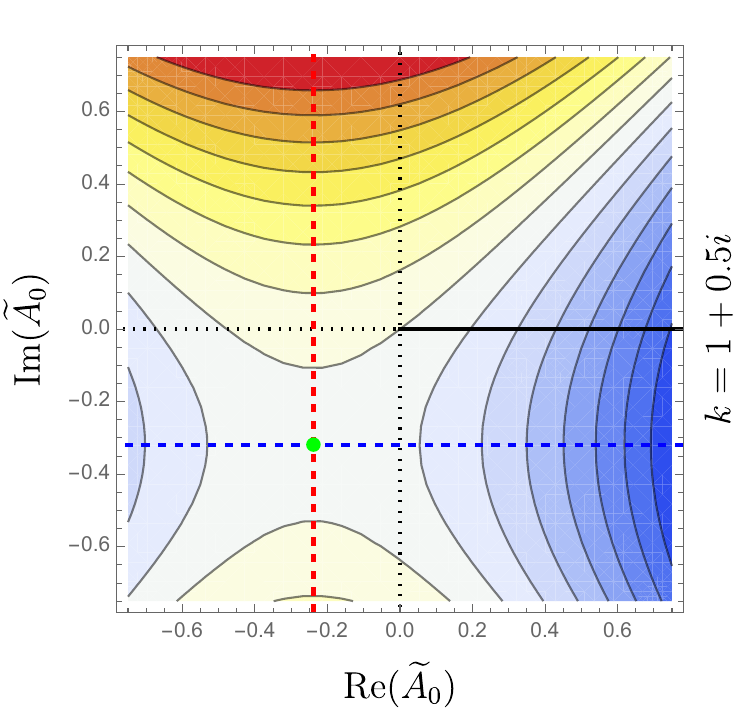}
\end{minipage}
\begin{minipage}{0.375\textwidth}
        \includegraphics[width=0.8\linewidth]{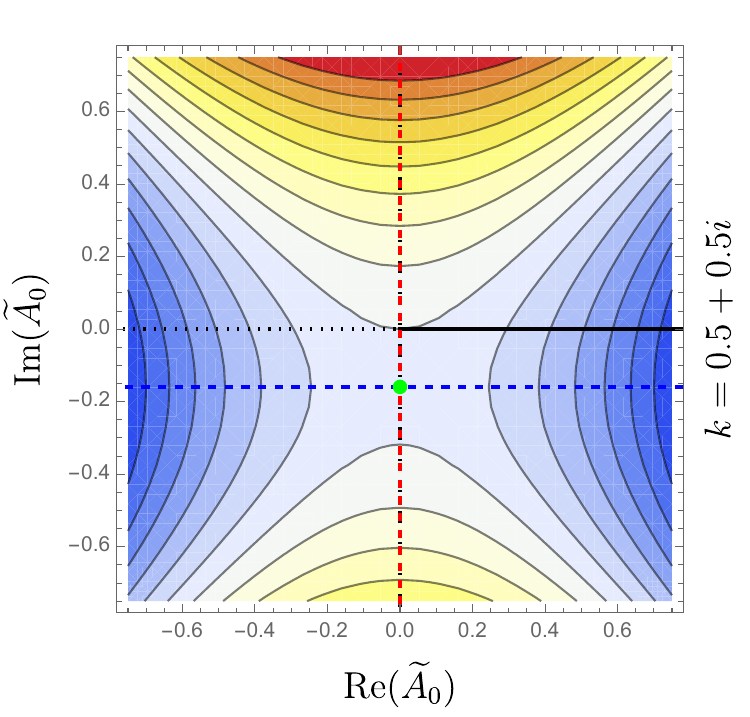}
\end{minipage}
\begin{minipage}{0.375\textwidth}
        \includegraphics[width=0.8\linewidth]{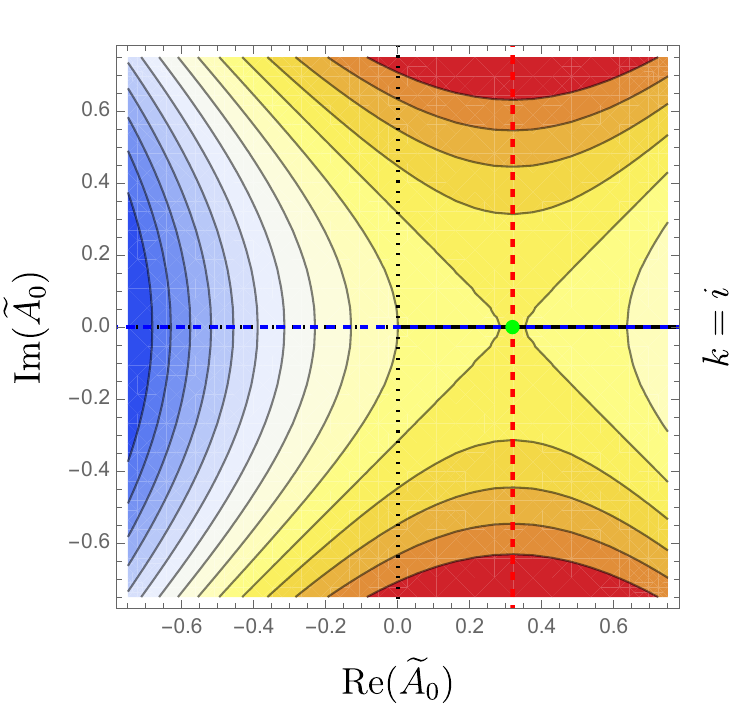}
\end{minipage}

\caption{A sequence of contour plots showing $\Re(iS_{CWH})$ in the complex $\tilde A_0$ plane for various $k\in \mathbb C$ and for fixed parameters $\beta=1$, $\phi_b=10^3$. Warmer colors indicate larger $\Re(iS_{CWH})$ while cooler colors indicate smaller values. Green dots mark the saddle $\tilde A_0=-\frac{k^2}{\pi \phi_b}$, and the blue and red dashed lines are the steepest descent and ascent contours of the saddle respectively. The solid black line is the integration contour, while the  real and imaginary axes are shown using dotted black lines.}
\label{fig:AreaContour}
\end{figure}
However, the location of the saddle $\tilde A_0=-\frac{k^2}{\pi \phi_b}$ depends on $k$ as shown in figure \ref{fig:AreaContour}. Due to the simple form of the action, there are simple analytic expressions for the associated steepest descent and ascent contours:
\begin{eqnarray}
      \text{steepest descent:}\quad  \Im(\tilde A_0)= -\frac{2\Re(k)\Im(k)}{\pi\phi_b},\\
      \text{steepest ascent:}\quad  \Re(\tilde A_0)= \frac{\Im(k)^2-\Re(k)^2}{\pi\phi_b}.
\end{eqnarray}

For general $k\in\mathbb C$, the saddle point contributes to the path integral when the steepest ascent contour intersects the integration contour $\tilde A_0 \in [0,+\infty)$. We can see from the above formula that this happens when $|\Im(k)|>|\Re(k)|$, in which case the integral evaluates to $Z(\beta) = \exp \left(\frac{k^4\beta}{\pi^2\phi_b} \right)$. When $|\Im(k)|<|\Re(k)|$, the integral remains dominated by the endpoint contribution from $\tilde A_0=0$, so again $Z(\beta)=O(1)$. For the marginal case $|\Im(k)|=|\Re(k)|$, the saddle lies on a Stokes' ray.   There is thus a choice as to how one wishes to define the endpoint contribution and this choice determines whether one takes the saddle to give a separate contribution.  However, even when one takes the saddle to contribute, it is subdominant to the endpoint contribution.

In particular, when $k$ becomes purely imaginary, we see that the saddle lies on the contour of integration. As a result, the wormhole saddle contributes to the Lorentzian gravitational path integral for all imaginary $k$.

\section{Discussion}
\label{sec:disc}

This work used Lorentzian methods to study the relevance of the imaginary-scalar JT wormhole \cite{Garcia-Garcia:2020ttf} to the two-partition function at imaginary boundary values of the scalar field.  The analysis was in direct parallel to our recent study \cite{AxionWormholes} of AdS axion wormholes, though with significant simplifications.  Nevertheless, the results are quite different.  

In particular, while neither wormhole is relevant at real boundary values of the scalar/axion\footnote{Using a convention analogous to the one used here in which the axion kinetic term has the usual sign even in Euclidean signature.}, even at imaginary axion  ref. \cite{AxionWormholes} found that AdS-axion wormholes provide at most a subdominant contribution.    In contrast, in the JT case we found the wormhole to become relevant more quickly as we analytically continued the scalar boundary values away from the real line, and for imaginary boundary values we found the wormhole of \cite{Garcia-Garcia:2020ttf} to dominate the integral.  This supports the interpretation suggested in \cite{Garcia-Garcia:2020ttf} that the wormhole describes a bulk dual of the physics associated with ensembles of SYK systems with imaginary couplings.

We thus see that the relevance of wormhole saddles can depend on details of the system studied.  While there are many similarities between the current JT system and the AdS$_3$ axion system studied in \cite{AxionWormholes}, there are also important differences.  In particular, we note that JT gravity describes a dimensional reduction of Einstein gravity deep in the throat of extremal black holes, but that it is certainly not an exact description of Einstein gravity in the uncharged context analyzed in \cite{AxionWormholes}.  

It might thus be of interest to perform a similar study of charged axion wormholes and to look for phase transitions as one changes the total charge.  It would also be useful to analyze other classes of wormholes in the same way (and in particular to do so for various wormholes that were found to be stable in \cite{Marolf:2021kjc}).

Returning to the JT-scalar system, the fact that fixing $\tilde A_0$ provides constrained-saddles on the original contour of integration distinguishes $\tilde A_0$ from generic degrees of freedom and to some extent justifies the importance of treating this particular mode in detail.  Nevertheless, 
one would of course like to generalize our study to consider other degrees of freedom in detail as well.  In particular, for real $\chi_\infty$ and real $T$ we used the fact that our area-constrained wormhole lies on the real Lorentzian contour to justify its relevance to our path integral at fixed $\tilde A_0$ and fixed real $T$ (see e.g. footnote \ref{foot:definingC}).  But our evaluation of the $T$-integral using the pole at $T=-i\beta$ involves analytic continuation into the complex $T$-plane.  There might thus in principle be Stokes' phenomena that intervene and change the relevance of our fixed-$\tilde A_0$ saddle as a function of $T$.  Tracking the details of additional degrees of freedom could shed more light on this issue.    

It is also important to emphasize that our analysis here worked entirely at leading semiclassical order.  In contrast, it was argued in \cite{Garcia-Garcia:2020ttf} that (as usual in JT gravity) quantum corrections are important at low temperatures. It would thus be useful to generalize the present analysis to include 
such effects.

However, the most important open question may be to understand if our Lorentzian approach defines a general recipe that can be applied directly in Euclidean signature.  For example, in the current context, if we first decide to take the scalar to be purely imaginary, a Euclidean analysis of fixed-area constrained wormholes would simply reproduce the results of the final panel in figure \ref{fig:AreaContour} along the real $\tilde A_0$-axis.  This might again be taken to suggest that the saddle dominates.   More generally, one would like to better understand the relation to various perturbative frameworks that can provide more practical ways to simultaneously analyze infinite families of degrees of freedom.

\acknowledgments  We thank Xiaoyi Liu for helpful discussions. We also thank Douglas Stanford for bringing this wormhole to our attention and for emphasizing its physical interpretation.  JH was supported on this project by U.S. Department of Defense through the National Defense Science and Engineering Graduate (NDSEG) Fellowship Program as well as the U.S. Air Force Office of Scientific Research under award number FA9550-19-1-0360, by NSF grant PHY-2408110, and by funds from the
    University of California.   DM was supported by NSF grant PHY-2408110 and by funds from the University of California.  MK was supported by NSF grant PHY-2107939, by funds from the University of California, by the European Research Council (ERC) under the European Union’s Horizon 2020 research and innovation program (grant agreement No. 884762), and by the ERC under the QFT.zip project (grant agreement No. 101040260). ZW was supported by the DOE award number DE-SC0015655.

\addcontentsline{toc}{section}{References}
\bibliographystyle{JHEP}
\bibliography{ref.bib}
\end{document}